\documentclass[12pt]{article}
\usepackage{graphics}
\let\a=\alpha \let\b=\beta    \let\d=\delta \let\e=\varepsilon
\let\z=\zeta     \let\th=\theta  
\let\m=\mu    \let\n=\nu         \let\p=\pi    \let\r=\rho
\let\s=\sigma     
   
 \let\D=\Delta  \let\L=\Lambda

\def\\{\hfill\break} \let\==\equiv

\let\0=\noindent

\def\nn{\nonumber}

\def\qed{\hfill\raise1pt\hbox{\vrule height5pt width5pt depth0pt}}

\def\be{\begin{equation}}
\def\ee{\end{equation}}
\def\bea{\begin{eqnarray}}\def\eea{\end{eqnarray}}
\renewcommand{\theequation}{\arabic{section}.\arabic{equation}}
\begin{document}
%------------------------------------------------------------------------------
\markright{Stability Limit ...}
%------------------------------------------------------------------------------

\title{Absolute Stability Limit for Relativistic Charged Spheres}

\author{Alessandro Giuliani{\small\it \thanks{E-mail: giuliani@princeton.edu.
On leave from Dipartimento di Matematica, Universit\'a di RomaTre,
Largo San Leonardo Murialdo 1, 00146, Roma, Italy.}} ~and Tony
Rothman{\small\it\thanks{E-mail: trothman@princeton.edu.}}
\\[2mm]
%{\small\it \thanks{giuliani@princeton.edu}}
%\\
%{\small\it\thanks{trothman@princeton.edu}}
~ \it Princeton University, Princeton NJ 08540}

\date{{\small   \LaTeX-ed \today}}
%-----------------------------------------------------------------------------

\maketitle

%-----------------------------------------------------------------------------
\begin{abstract}
We find an exact solution for the stability limit of relativistic
charged spheres for the case of constant gravitational mass density
and constant charge density. We argue that this provides an absolute
stability limit for any relativistic charged sphere in which the
gravitational mass density decreases with radius and the charge
density increases with radius.  We then provide a cruder absolute
stability limit that applies to any charged sphere with a
spherically symmetric mass and charge distribution. We give
numerical results for all cases. In addition, we discuss the example
of a neutral sphere surrounded by a thin, charged shell.

 \vspace*{5mm} \noindent PACS: 04.70, 04.70 Bw,
97.60.Lf,
\\ Keywords: Extremal Black holes, Reissner-Nordstr\" om solution,
Stellar Stability, Oppenheimer-Volkov Equation.
\end{abstract}

%-----------------------------------------------------------------------------
\section{Introduction}
\setcounter{equation}{0}\label{sec1}
%-----------------------------------------------------------------------------
\baselineskip 8mm

Over the past decade, extremal black holes---black holes for which
the charge equals the mass in geometric units---have been the
subject of considerable interest, largely because such objects were
the ones originally employed to derive the Bekenstein-Hawking
entropy directly from string theory\cite{Strom96}.  Aside from
developments in string theory, however,  there has long been ample
motivation to study extremal black holes  because from the classical
and semi-classical point of view they provide the
``zero-temperature'' limit in black hole thermodynamics. At the same
time,  substantial  evidence suggests that one should not view
extremal black holes as any sort of continuous limit of their
sub-extremal counterparts, black holes for which the charge is less
than the mass. For example, the horizon structure of classical,
charged black holes changes completely at
extremality\cite{Hawking73}. Some studies have also concluded that
entropy is not well-defined for extremal black
holes\cite{Das97,LRS00}.  More definitely, one knows from Israel's
proof of the third law of black hole dynamics\cite{Israel86} that
extremality cannot be attained in a finite time, and that the
conclusion holds even under Hawking radiation and superradiance,
which violate the assumptions of Israel's proof\cite{Das97,AR02}.
Thus, it appears impossible to create an extremal black hole from a
subextremal one, and the only remaining possibility is to produce
one from the collapse of an already extremal object.

For this reason it is of interest to investigate the stability of
relativistic charged spheres. Previous studies along these lines
have been mainly numerical\cite{deF99,AR02} and have concluded that
while for $Q<M$ collapse always takes place at a critical radius
$R_c$ outside the horizon, as $Q$ approaches $M$, this critical
radius approaches the horizon itself, $R_+$. The present paper is
intended as an analytic companion to the numerical investigations.
Our point of departure is the classic proof of
Buchdahl\cite{Buchdahl59}, who showed that for uncharged stars
gravitational collapse into a black hole will always take place when
$R < (9/4) M$, regardless of equation of state.\footnote{Throughout
we use units in which $G = c = 1$.} (See Weinberg's {\it Gravitation
and Cosmology}\cite{Weinberg72} for a clear presentation of
Buchdahl's argument, or \S \ref{sec4} of this paper.) Because in a
charged sphere Coulomb repulsion tends to oppose the gravitational
force, $R_c$  should be less than $(9/4)M$. However,
relativistically, the charge increases gravitational energy as well,
and so at some point gravity always wins out and collapse into a
black hole takes place.  At $Q=M$, the Coulomb repulsion equals the
gravitational force and one finds numerically that  $R_c = R_+$.
Thus one should have $M \le R_c < (9/4)M$, always.

Although for a given charge distribution one can indeed find $R_c$
numerically, one does suspect that there must be an analytic proof,
analogous to Buchdahl's, that applies to relativistic charged
spheres.\footnote{In this paper we tend to speak of charged spheres
rather than charged stars, as there is no good reason to think that
charged stars, in the usual sense of the word, exist.} In other
words, given a value of $Q/M$ we should be able to find an absolute
bound on $R/M$, independent of other physical parameters, below
which the object collapses into a black hole. Anninos and
Rothman\cite{AR02} (henceforth AR) intended to include such a proof
as a supplement to their numerical investigation but as that project
neared completion they learned that Yunqiang and Siming (henceforth
YS) had already claimed to have given such a proof\cite{Yunqiang99}.
The YS demonstration, however, is far from transparent and does not
appear to have ever been published. Moreover it does not provide a
sharp value for the collapse radius, as in Buchdahl's proof, but
rather gives a general lower bound on it. The important feature of
this lower bound is that for $Q<M$ it is {\it always  larger than
$R_+$}, as expected from numerics. Not long ago we decided to take
the opportunity to present a simplified version of this interesting
result.  In the process we have found an exact solution for the case
of constant mass and charge densities, and this allows us, in a
manner complete analogous to Buchdahl's, to put an absolute (sharp)
stability limit on a very large class of objects, all those with
charge density increasing radially, and gravitational mass density
decreasing radially.  This stability limit, which should cover
essentially all cases of interest, is the main result of our paper.
For the remaining cases, we present a proof similar to that of YS,
but we hope with greater clarity, and give explicit numerical
results for a lower bound on $R_c$. We also give an exact solution
for the stability limit of a neutral sphere surrounded by a charged
shell.

The paper is organized as follows. In Section \ref{sec2} we
introduce the relevant Einstein equations and introduce notation. In
Section \ref{sec4} we review the $Q=0$ case and derive an exact
solution for the case of a neutral sphere surrounded by a charged
shell. Section \ref{sec5} is devoted to the main result of our
paper: we solve exactly the case of constant charge density and
constant gravitational mass density, derive its critical stability
radius and show that it gives an absolute bound on the critical
stability radii for all spherically symmetric distributions in which
gravitational mass density gradient is negative and the charge
density gradient is positive. In Section \ref{sec7} we calculate a
general lower bound on the critical stability radius. Finally in
Section \ref{sec8} we summarize the results and draw conclusions.

%-----------------------------------------------------------------------------
\section{Einstein Equations}
\setcounter{equation}{0}\label{sec2}
%-----------------------------------------------------------------------------

As mentioned above, the plan is to find an absolute stability limit
on $R/M$ for relativistic charged spheres that is independent of the
equation of state and
 depends only on $Q/M$. We will assume throughout
that the pressure $p$ and density $\rho$ are both positive, that the
charge density is positive and that $Q\le M$  in order to avoid
naked singularities; this last assumption ensures that spacetime is
asymptotically predictable\cite{Hawking73}.

We also restrict attention to spherically symmetric mass and charge
distributions, for which the metric can be written in the form
\begin{equation}
ds^2 = -e^{2\Phi(r)}{\rm d}t^2 + e^{2\Lambda(r)}dr^2 + r^2{\rm
d}\theta^2 + r^2{\rm sin^2}\theta d\phi^2, \label{metric}
\end{equation}
where the metric components $e^{2\Phi(r)}$ and $e^{2\Lambda(r)}$ are
positive.

As is well-known, the classic Reissner-Nordstr\"om (RN) solution for
the charged spherically symmetric case gives
\begin{equation}
e^{-2\Lambda(r)} = 1 - \frac{2 M}{r} + \frac{Q^2}{r^2} =
e^{+2\Phi(r)} \label{lambdamatch},
\end{equation}
where $r\ge R$ and $R$ is the outer radius of the sphere.  The
Schwarzschild-Droste (SD) solution is of course recovered by setting
$Q = 0$.  Both the RN and SD, however, are vacuum solutions,
concerned solely with the metric outside $R$. For the collapse
problem we need to study the behavior of the metric functions
$\Lambda(r)$ and $\Phi(r)$ for $r < R$, where the pressure, the
charge and mass densities are nonzero. The procedure for solving the
``interior Reissner-Nordstr\"om equations" is nevertheless much the
same as for the exterior case. One assumes (see AR or de
Felice\cite{deF99} for more details) a perfect-fluid stress-energy
tensor for the hydrodynamic part
\begin{equation}
(T_{\mu\nu})_{hydro} = (p+\rho)u_\mu u_\nu + pg_{\mu\nu},
\end{equation}
while for the electromagnetic part
\begin{equation}
4\pi (T_{\mu\nu})_{EM} = F_\mu^\alpha F_{\nu\alpha} -
\frac{1}{4}g_{\mu\nu}F_{\alpha\beta}F^{\alpha\beta} =
\frac{q^2(r)}{8\pi r^4}\;\rm {diag} [~e^{2\Phi},\; -e^{2\Lambda},\;
r^2,\; r^2\rm{sin}^2\theta~].\label{TEM2}
\end{equation}
In these expressions $\r=\r_{rm}+e$ is the total mass density,
$\r_{rm}$ is the rest mass density, $e$ is the internal energy
density, $p$ is the fluid pressure, $u_\m$ is the four-velocity and
$F_{\mu\n}$ is the electromagnetic field strength tensor.
$(T_{\mu\nu})_{EM}$ here is of the same form as for the exterior RN
solution, as it must be by Gauss's law, except that instead of the
total charge $Q$, we now have a $q(r)$, the charge within any given
radius $r$. Indeed, by definition
\begin{equation}
q(r) = 4\pi \int_0^r e^{\Phi(r') + \Lambda(r')}r'^2 j^0(r')dr'
\label{q},
\end{equation}
where $j^0$ is the charge density.  (This is the usual definition of
charge, modified only for metric curvature.) The boundary condition
requires $Q = q(R)$.

With the above stress-energy tensor, the (00) Einstein equation is
found to be
\begin{equation}
\Phi'' + \Phi'^2 -\Phi'\Lambda' + \frac{2\Phi'}{r}
 = 4\pi e^{2\Lambda}\left[\rho + 3p + \frac{q^2(r)}{4\pi
 r^4}\right],\label{00}
\end{equation}
where $\prime$ denotes derivatives with respect to $r$. Similarly,
the (11) equation is
\begin{equation}
 -\Phi'' - \Phi'^2
+\Phi'\Lambda' + \frac{2\Lambda'}{r}
 = 4\pi e^{2\Lambda}\left[\rho - p - \frac{q^2(r)}{4\pi r^4} \right],
 \label{11}
\end{equation}
and the (22) equation is
\begin{equation}
e^{2\L} -1 + \L' r -\Phi' r = 4\pi r^2e^{2\L}\left[\rho - p +
\frac{q^2(r)}{4\pi r^4} \right]\label{22}.
\end{equation}

The left-hand-side of these equations is necessarily the same as for
the exterior SD or RN solutions; only the right-hand-side differs
because of the nonzero stress-energy tensor. Following the standard
procedure for deriving the exterior solutions, we can take linear
combinations of Eqs. (\ref{00}), (\ref{11}) and (\ref{22}) to
eliminate the terms in $\Phi$. One easily finds that for any $r\le
R$
 \begin{equation}
e^{-2\Lambda(r)} = 1 - \frac{2 m_i(r)}{r}-
 \frac{{\cal F}(r)}{r},\label{eL}
\end{equation}
where
\begin{equation}
 m_i(r) \equiv 4\pi\int_0^r\rho\; r'^2\; dr' \ \  \ \ \mbox{and}\ \
 \ \ {\cal F}(r) \equiv \int_0^r\frac{q^2(r')}{r'^2}\;
 dr'. \label{mf}
 \end{equation}
Here, $m_i(r)$ is the usual definition of the mass within a radius
$r$. The subscript {\it i} denotes ``internal" to emphasize that
$m_i$  contains both rest and internal energy. We use the
designation because it will become necessary  to distinguish
$m_i(r)$ from the gravitational mass $m_g(r)$, defined momentarily.
Requiring that  $(\ref{eL})$ matches the exterior solution
(\ref{lambdamatch}) at $r=R$, gives
\begin{equation}
1 - \frac{2M}{R} + \frac{Q^2}{R^2}
 = 1-\frac{1}{R}\int_0^R(8\pi\rho r^2 +
 \frac{q^2}{r^2})dr, \label{match}
\end{equation}
or
\begin{equation}
M = \frac{1}{2}\int_0^R( 8\pi\rho r^2 + \frac{q^2}{r^2})dr +
\frac{Q^2}{2R}, \label{gravmass}
\end{equation}
which defines the  gravitational mass at $R$ (the mass measured by a
satellite in orbit around the object).  By Gauss's law, however, the
same must be true at any radius, and so using the definition of
$\cal F$ from Eq. (\ref{mf}),
\begin{equation}
m_g(r) =  m_i(r) + \frac{{\cal F}(r)}{2} + \frac{q^2(r)}{2r},
\label{gravmassr}
\end{equation}
which clarifies the distinction between $m_i$ and $m_g$.  In terms
of the gravitational mass, the metric function $e^{\Lambda(r)}$ is
\begin{equation}
e^{\Lambda(r)} = \left(1 - \frac{2 m_i(r)}{r}-
 \frac{{\cal F}(r)}{r}\right)^{-1/2}  = \left(1 - \frac{2 m_g(r)}{r}
 + \frac{q^2(r)}{r^2}\right)^{-1/2} .\label{Lg}
\end{equation}
One can write these functions either in terms of $m_i$ and $\cal F$,
or $m_g$ and $q$, but because we do not in general know the charge
distribution $q(r)$ and hence ${\cal F}(r)$,  when thinking about
boundary conditions it is much more convenient to use $m_g$, since
in that case $e^{\Lambda(r)}$ matches onto $e^{\Lambda(R)}$ in the
expected way. We will therefore  generally use the second form.

The pressure in these equations can be eliminated by taking three
times Eq. (\ref{11}) and adding it to  Eq. (\ref{00}), which yields
\[
\Phi'' + \Phi'^2 -\Phi'\Lambda' - \frac{\Phi'}{r}
 = \frac{3 \Lambda'}{r} - \left(8\pi \rho  - \frac{q^2}{r^4}\right)
e^{2\Lambda}.
\]
Upon multiplication by $e^{-\L+\Phi}/r$, the left-hand-side turns
out to be an exact differential, and so, letting $\zeta(r) \equiv
e^{\Phi(r)} $ as in Weinberg's notation \cite{Weinberg72}, one has
\begin{equation}
\left(\frac{1}{r} e^{-\Lambda}\zeta'\right)' =
\left[\frac{3\Lambda'e^{-2\Lambda}}{r^2} - \frac{8\pi\rho}{r} +
\frac{q^2}{r^5}\right]e^\Lambda\zeta .\label{zeta}
\end{equation}

Eq. (\ref{zeta}) will prove to be the fundamental equation of our
analysis. It can be brought into perhaps more familiar form by
noting by that $8\pi\rho/r = 2m_i'(r)/r^3$ and ${\cal F}'(r) =
q^2/r^2$. Then
\begin{equation}
\left(\frac{1}{r} e^{-\Lambda}\zeta'\right)' =e^{\L(r)}
%\left(1-\frac{2m_i(r)}{r} - \frac{{\cal F}(r)}{r}\right)^{-1/2}
\left[\left(\frac{m_i(r)}{r^3}\right)' + \frac{1}{2r^3}\left(\frac{5
q^2(r)}{r^2} - \frac{3{\cal F}(r)}{r}\right) \right]\zeta.
\label{zetadiff}
\end{equation}
This is the equivalent of Weinberg's Eq.(11.6.14), employed in
Buchdahl's proof of the absolute limit of stability for ordinary
stars. Our equation, however, contains two more terms within the
square brackets than the usual one, as well as an extra term in the
expression (\ref{Lg}) defining $e^{\L}$. We point out that although
$e^{\Lambda(r)}$ has the simple form given in Eq.~(\ref{Lg}), no
such closed form exists for $\zeta(r)$ for $r\le R$. Indeed the
differential equation (\ref{zeta}) should be regarded as the
equation defining $\zeta$ in the interior of the sphere {\it for
given input distributions} $\r$ and $q$.

Requiring that $\z$ and $\z'$ match on to the exterior RN solution
at $r=R$ gives the following important boundary conditions:
\begin{eqnarray}
 \zeta(R)
            &=&\left(1-\frac{2M}{R}+\frac{Q^2}{R^2}\right)^{1/2}\nonumber\\
\zeta'(R) &= &\left(1-\frac{2M}{R}+\frac{Q^2}{R^2}\right)^{-1/2}
            \left(\frac{M}{R^2} -
            \frac{Q^2}{R^3}\right)\;,\label{zetamatch}
\end{eqnarray}
with $M$ the gravitational mass in (\ref{gravmass}).

Once $\z$ and $\Phi$ have been computed using (\ref{zeta}), one can
easily show from (\ref{22}), (\ref{eL}) and (\ref{gravmassr})
 that the pressure is given in terms of $\Phi'$, $m_g$ and $q$ by
\be p=\frac1{4\p
r^2}\left(\Phi're^{-2\L}-\frac{m_g}{r}+\frac{q^2}{r^2}\right).
\label{press}\ee
Clearly, in order for the solution of (\ref{zetadiff}) to be
physically acceptable, we must require that $\z>0$ and that $p$, as
computed from (\ref{press}), is nonnegative and satisfies a proper
equation of state, usually assumed of the form
$p=p\big[\r_{rm}\big]$. For a given class of distributions $\r,q$ we
define the critical stability radius $R_c$ as the smallest possible
radius for which a physically acceptable solution to (\ref{zeta})
can be found.

Then, in order to find a bound on the stability of the charged star,
it is sufficient to show that if the radius is smaller than $R_c$,
then no physically acceptable solution exists, whatever  the choice
of $\r,q$.

%-----------------------------------------------------------------------------
\section{Neutral Sphere Surrounded by Charged Shell}
\setcounter{equation}{0}\label{sec4}
%-----------------------------------------------------------------------------

In this section we first review Buchdahl's $Q = 0$ case, which
illustrates the general strategy for finding $R_c$.  The question
then naturally arises as to whether the result  changes for  a
neutral sphere surrounded by a charged shell. We show that the basic
$Q=0$ scenario is fairly easily adapted to cover this case.

For the moment, then, let us set $Q = 0$ and drop the subscript $i$
on $m$ (there is now no distinction between between $m_g$ and
$m_i$). Eq.  (\ref{zetadiff}) then becomes
\begin{equation}
\left(\frac{1}{r} e^{-\Lambda}\zeta'\right)' =\left(1
-\frac{2m(r)}{r}\right)^{-1/2}\left[\left(\frac{m(r)}{r^3}\right)'\right]\zeta,
\label{zetadiff0}
\end{equation}
where we have used   $e^{-2\Lambda(r)} = 1 - {2 m(r)}/{r}$.

Let us now assume that for any physically reasonable star $\r'\le
0$. Then, because $(m/r^3)'=4\p r^{-4}\int_0^r\r'(x)x^3dx\le 0$,  it
follows that
\begin{equation}
\left(\frac{1}{r} (1-\frac{2m}{r})^{1/2}\zeta'\right)' \le
0\label{zetadiff00}
\end{equation}
Integrating this expression from $r$ to $R$ gives
\begin{equation}
\frac{1}{R}
\left(1-\frac{2M}{R}\right)^{1/2}\zeta'(R)-\frac{1}{r}\left(1-\frac{2m}{r}
\right)^{1/2}\zeta'(r) \le 0. \label{zeta'0}
\end{equation}

We now make use of the boundary conditions Eq. (\ref{zetamatch})
with $Q=0$. Inserting the expression for $\zeta'(R)$ into Eq.
(\ref{zeta'0}) gives
\begin{equation}
\zeta'(r) \ge \frac{Mr}{R^3}\left(1-\frac{2m(r)}{r}\right)^{-1/2}.
\end{equation}
Integrating again from 0 to $R$ yields
\begin{equation}
\zeta(R) - \zeta(0) \ge \frac{M}{R^3}\int_0^R\frac{r\,
dr}{\left(1-\frac{2m(r)}{r}\right)^{1/2}}.\label{zetaR}
\end{equation}
with $\z(R)=\left(1-\frac{2M}{R}\right)^{1/2}$. Now, in order to
have $\zeta(0)>0$, we require that
\begin{equation}
0 < \zeta(0) \le \left(1-\frac{2M}{R}\right)^{1/2} -
\frac{M}{R^3}\int_0^R \frac{r\,
dr}{\left(1-\frac{2m(r)}{r}\right)^{1/2}}.\label{in}
\end{equation}
Note that, as remarked above, $(m/r^3)'\le0$, and so $m/r^3\ge
M/R^3$ for all $r\le R$. Plugging this into (\ref{in}) we find:
\begin{equation}
0 <\left(1-\frac{2M}{R}\right)^{1/2} -
\frac{M}{R^3}\int_0^R\frac{r\,
dr}{\left(1-\frac{2Mr^2}{R^3}\right)^{1/2}}.\label{zeta0}
\end{equation}
The integral is now trivially performed to get \be 0 <
\left(1-\frac{2M}{R}\right)^{1/2}
-\frac1{2}\left[1-\left(1-\frac{2M}{R}\right)^{1/2}\right],
\label{zeta0Q} \ee which immediately implies Buchdahl's result $R >
(9/4)M$.

Note that for stars with constant density all the above inequalities
become equalities at the critical radius, and so the value $(9/4)M$
is precisely their critical stability radius.\footnote{It is
straightforward to check that in the uncharged case, if $\z>0$, then
$p>0$ and finite, $\forall r<R$. This means that the only condition
to be imposed for the solution to be physically acceptable is
$\z>0$, in other words, precisely the condition we imposed above.}
The value $R_c =(9/4)M$ gives an absolute (sharp!) stability limit
for all stars with distributions satisfying $\r'\le 0$. If any  such
star is compressed to the point
that $R <  R_c$ gravitational collapse necessarily takes place.\\

Let us now modify the above computation to handle the case of a
neutral sphere of constant density surrounded by a thin shell of
internal (``inertial") mass $K$ that carries a uniformly distributed
charge $Q\le M$. In such a situation,  Eq. (\ref{Lg}) shows that
$\L$ suffers a discontinuity at $r=R$.  Let $R^-$ and $R^+$
represent the inner and outer radii of the shell, $M_{int}$ be the
mass interior to the shell and  $M_s = K + Q^2/2R$ be the
gravitational mass of the shell (cf. Eq. (\ref{gravmass})).  Then $M
= M_{int} + M_s$ is the total mass and $e^{-2\L(R^-)} =
1-2M_{int}/R$, while $e^{-2\L(R^+)} = 1 - 2M/R + Q^2/R^2$.   Since
the jump in $\L$ is finite, however, Eq. (\ref{press}) implies that
any discontinuity in $\Phi'$ and hence in $\z'$ is finite as well.
Thus $\z$ itself is continuous at the boundary with precisely the
value given by the first of Eqs. (\ref{zetamatch}).

One can greatly simplify the calculations by assuming that $K = 0$,
in which case $\L$ and $\Phi$ are both {\it continuous} at $r=R$ and
only their derivatives suffer a discontinuity at the surface.  (For
a more detailed discussion of these issues we refer the reader to
Cohen and Cohen \cite{CC69}, who derive the solution for  a thin
charged shell of radius $R$, with $M_{int} = 0$.)

So, let us take $K=0$ and assume that interior to the shell
$\r=3M_{int}/(4\p R^3)$, with $M_{int}=M-Q^2/(2R)$. In order for the
  shell to be stable against gravitational collapse, it
is necessary to have a nonzero elastic stress tensor concentrated on
the surface, as assumed in \cite{CC69}. This means that the stress
energy tensor must be modified by the addition of a term
$(T_{\m\n})_{el}$ whose only nonzero elements are
$T_{\th\th}=r^2S\d(r-R)$ and $T_{\phi\phi}=r^2\sin^2\th S\d(r-R)$,
where $S$ is the elastic energy and the delta function is normalized
such that $\int dr\, 4\p r^2 \d(r-R)=1$. The presence of
$(T_{\m\n})_{el}$ modifies the Einstein equations as follows: Eq.
(\ref{00}) contains an extra term $8\p e^{2\L} S \d(r-R)$ on the
right hand side; Eq. (\ref{11}) contains an extra term $-8\p e^{2\L}
S \d(r-R)$ on the right hand side; Eq. (\ref{22}) is unchanged. With
these additions Eq. (\ref{zetadiff}) becomes
\begin{equation}
\left(\frac{1}{r} e^{-\Lambda}\zeta'\right)' =e^{\L(r)}
%\left(1-\frac{2m_i(r)}{r} - \frac{{\cal F}(r)}{r}\right)^{-1/2}
\left[\left(\frac{m_i(r)}{r^3}\right)' + \frac{1}{2r^3}\left(\frac{5
q^2(r)}{r^2} - \frac{3{\cal F}(r)}{r}\right)+ \frac{8\p}{r} S
\d(r-R)\right] \zeta,\label{zetadiffs}
\end{equation}
where
\be e^{-2\L(r)}=\cases{ 1-2m_i(r)/r & $r<R$ \cr 1-2M/r+Q^2/r^2 &
$r>R$}\ee
and $\z=e^{-\L}$ for $r\ge R$.  Integrating both sides of Eq.
(\ref{zetadiffs}) between  $R^-$ and $R^+$, and using the fact that
$\L$ and $\Phi$ are both continuous at $r=R$, we see that
$e^{-\L(R)}\big(\z'(R^+)-\z'(R^-)\big)= 2\z e^{\L}S/R^2$, or
\be
\z'(R^-)=\frac1{\sqrt{1-2M_{int}/R}}\left[\frac{M_{int}}{R^2}-\frac{Q^2}{2R^3}-
\frac{2S}{R^2}\right]\;.\ee
 Computing $\z'(R^-)$ from Eq. (\ref{press}) yields
\be
\z'(R^-)=\frac1{\sqrt{1-2M_{int}/R}}\left[\frac{M_{int}}{R^2}+4\p R
p_-\right],\ee
and comparing the two expressions shows that
\be 4\p R p_-=-\frac{Q^2}{2R^3}-\frac{2S}{R^2}.\ee
The parameter $S$ should be chosen such that $p_- \ge 0$.

We can now proceed as in the $Q=0$ case, integrating
$\left(\frac{1}{r} e^{-\Lambda}\zeta'\right)'=0$ from $0$ to $R^-$
with the new
 boundary conditions $\z(R^-)=\sqrt{1-2M_{int}/R}$ and $\z'(R^-)=\frac1{
\sqrt{1-2M_{int}/R}}\left[\frac{M_{int}}{R^2}+4\p R p_-\right]$. The
result is that the new critical radius is smallest when $p_-=0$, in
which case $S = -Q^2/4R$, consistent with the result in \cite{CC69}
for  $K= 0$.  We then find that  $R_c$ is precisely $(9/4)M_{int}$,
as one might expect from Gauss's law. With $M_{int}=M-Q^2/(2R_c)$,
solving for $R_c$ in terms of $M$ and $Q$ gives
\be
R_c=\frac98\left(M+\sqrt{M^2-\frac89Q^2}\;\right)\label{critR}.\ee
For $Q= 0$,  $R_c = 9/4M$, as expected, and a nonzero $Q$ indeed
lowers $R_c$.  The extremal case, $Q=M$, gives $R_c = 3/2M$, which
is plausible, as below we will find that for the full extremal
charged sphere $R_c = M$.

%-----------------------------------------------------------------------------
\section{Constant Density Case}
\setcounter{equation}{0}\label{sec5}
%-----------------------------------------------------------------------------

The special case of perhaps greatest interest (and the easiest one
to handle), is that of constant density, $m_g \propto r^3$ and $q
\propto r^3$. Remarkably, we are able to find an exact solution in
this situation.  Moreover, because a neutral test particle senses
the gravitational mass $m_g$ within a radius $r$, it is evidently
$m_g$ that plays the role $m_i$ did in the $Q=0$ case.  In other
words, it is $m_g$ that determines the weight of material in the
sphere, and a physically reasonably requirement for stability is
that $\rho_g'\le 0$, where $\r_g$ is the gravitational mass density.
If we additionally impose the requirement that the charge density is
positive and increases outwards (that is, $q'\ge 0$ and $(q/r^3)'\ge
0$), which also seems reasonable if like charges repel, then we will
also be able to find, in complete analogy with $Q=0$ case,  an
absolute lower bound on the critical radius of any charged sphere
meeting the two conditions.

We begin by rewriting the fundamental equation (\ref{zetadiff}) in
terms of $m_g$:
\be \left(\frac1{r}e^{-\L}\z'\right)'=\left(1 - \frac{2 m_g(r)}{r}
 + \frac{q^2(r)}{r^2}\right)^{-1/2}\left[\left(\frac{m_g}{r^3}\right)'-q
\left(\frac{q}{r^4}\right)'\right]\z\label{zetadiff2}\ee
With the {\it ansatz}  that $m_g=M(r/R)^3$ and $q=Q(r/R)^3$, the
first term in the square brackets vanishes and Eq. (\ref{zetadiff2})
becomes
\be \left(\frac1re^{-\L_c}\z'\right)'=e^{\L_c}\frac{Q^2
r}{R^6}\,\z\label{zpconst}, \ee
where now
\be
e^{-\L_c}=\sqrt{1-2\frac{M}R\left(\frac{r}{R}\right)^2+\frac{Q^2}{R^2}
\left(\frac{r}{R}\right)^4}\;.\label{lambdaconst} \ee
Here and in what follows the subscript $c$ refers to
``constant-density case."

Let us define a new variable $\widetilde \z$ such that
\be \widetilde\z(f_c(r))=\z(r)\ \ ;\ \ f_c(r) \equiv
R^{-2}\int_0^rdx\,x e^{\L_c(x)}\;. \label{ztilde}\ee
Substituting (\ref{ztilde}) into Eq. (\ref{zpconst}) gives at once
\be \frac{d^2\widetilde
\z}{df_c^2}=\frac{Q^2}{R^2}\widetilde\z\label{ztildediff},\ee
which has the obvious solution
\be \widetilde \z_c(f_c)=c_1e^{Qf_c/R}+c_2e^{-Qf_c/R}.\label{zcfc}
\ee

Moreover,  $f_c(r)$ is a standard integral:

\[
f_c(r)=\int_0^rdx\frac{x/R^2}{\sqrt{1-2\frac{M}R\left(\frac{x}{R}
\right)^2+\frac{Q^2}{R^2}\left(\frac{x}{R}\right)^4}}=
\frac{R}{2M}\int_0^{Mr^2/R^3}\frac{dy}{\sqrt{1-2y+(\frac{Q^2
}{M^2})y^2}}
\]
\be =-\frac{R}{2Q}\log\left(\frac{M}{Q}-\frac{Q}{M}y
+\sqrt{1-2y+(\frac{Q^2 }{M^2})y^2}\right)\Bigg|^{Mr^2/R^3}_0, \ee
or\\
\be f_c(r)
=\frac{R}{2Q}\log\frac{M/Q+1}{M/Q-Qr^2/R^3+e^{-\L_c(r)}}.\label{fr}
\ee\vspace{2mm}

Imposing the boundary conditions
$\widetilde\z_c(f_c(R))=e^{-\L_c(R)}$ and
$d\widetilde\z_c(f_c(R))/df_c =M/R-Q^2/R^2$ we find after some
algebra
\bea&&
c_1=\frac12\frac{(M/Q-Q/R+e^{-\L_c(R)})^{3/2}}{\sqrt{M/Q+1}}\nn\\
&&c_2=-\frac12\left(M/Q-Q/R-e^{-\L_c(R)}\right)\sqrt{
\frac{M/Q+1}{M/Q-Q/R+e^{-\L_c(R)}}} ,\label{5.4} \eea
and so, finally, the exact solution for $\z$ is
\bea \z_c(r)&=&\frac12 \Biggl\{
\frac{\Big(M/Q-Q/R+e^{-\L_c(R)}\Big)^{3/2}}{\sqrt{M/Q-Qr^2/R^3+e^{-\L_c(r)}}}
\,-\cr
&&-\Big(M/Q-Q/R-e^{-\L_c(R)}\Big)\sqrt{\frac{M/Q-Qr^2/R^3+e^{-\L_c(r)}}
{M/Q-Q/R+e^{-\L_c(R)}}}\Biggr\}\label{zex} \eea
The condition for this solution to be  physical is
$\z_c(0)>0$.\footnote{A straightforward computation shows that if
$\z_c>0$ then $p$, as computed from (\ref{press}), is automatically
positive and finite, as expected.} As in the $Q=0$ case we get the
equation for the critical radius by setting $\z_c(0) = 0$, which
yields:
\bea&& \Big(M/Q-Q/R+e^{-\L_c(R)}\Big)^2=\left({M}/Q+1\right)
\left(M/Q-Q/R-e^{-\L_c(R)}\right)
%=\nn\\
%&&=\Big(M/Q-Q/R-e^{-\L(R)}\Big)\sqrt{\frac
%{M/Q-Q/R-e^{-\L(R)}}{M/Q-1}}
\label{5.6}\eea
One easily sees from Eq. (\ref{5.6}) that $Q = M$ implies $R_c = M$,
as claimed in the Introduction.  For other values of $Q/M$ we  solve
this equation for $R/M$.  Letting  $\m=M/R$ and $\s = Q/M$ in Eq.
(\ref{5.6}), we find after some further algebra:
\be 4\s^4\m^3-12\s^2\m^2+(9+3\s^2)\m-4=0.\label{cubic2} \ee
Thus the exact solution for the critical radius boils down to
finding the roots of this cubic equation for $\mu$  in terms of $\s
$. One immediately sees that $Q = 0$ implies that $R_c/M = 9/4.$
Numerical results for various values of $Q/M$ are given in Table 1
and plotted in Figure 1.
\newpage
\begin{center}
\begin{tabular}{|c||c|}
\hline Q/M     &                 \hbox{$R_c$}/M\\

0      &                2.250\\

.1      &                    2.244\\

.2       &                  2.226\\

.3       &                  2.196\\

.4       &                  2.152\\

.5      &                 2.093\\

.6       &                  2.016\\

.7       &                   1.915\\

.8       &                  1.781\\

.9       &                  1.586\\

.99       &                  1.224\\

.999      &                1.091\\

.9999     &                1.039\\
\hline
\end{tabular}

\vspace{5mm}

Table 1.  The stability limit $R_c/M$ for the constant-density
sphere, tabulated for various values of $Q/M$.
\end{center}

\begin{figure}[htb]
\vbox{\hfil\scalebox{.7}
{\includegraphics{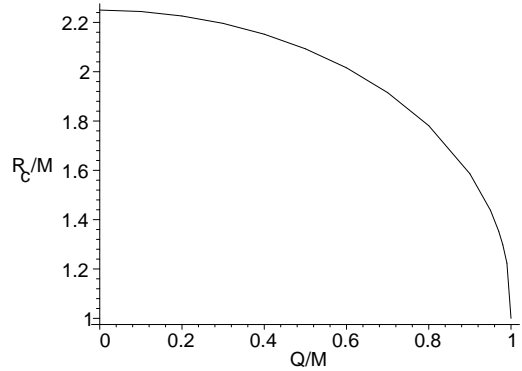}}\hfil}
{\caption{\footnotesize{The stability limit  $R_c/M$ plotted against
$Q/M$ for the constant density case.}}}
\end{figure}

Thus, we have found an exact solution for the constant density case.
We now argue, as indicated earlier, that any physically reasonable
solution should have $\r_g' \le 0$ and, for a positive charge
density, $(q/r^3)' \ge 0$.  If so, the constant density case
maximizes the expression in square brackets in the right-hand-side
of Eq. ({\ref{zetadiff2}}) and then $\left((1/r)e^{-\L}\z'\right)'
\le e^{\L}({Q^2r}/{R^6})\z$, as long as $\z\ge 0$.  In terms of
$\widetilde\z$ and $f=R^{-2}\int_0^rdx\,x e^{\L(x)}$, Eq.
(\ref{ztildediff}) is now replaced by the inequality
\be \frac{d^2\widetilde\z}{df^2}\le
\frac{Q^2}{R^2}\widetilde\z.\label{zineq}\ee
Note that now $f\neq f_c$, because the $e^{\L}$ appearing in the
definition of $f$ is no longer that of the constant-density solution
(\ref{lambdaconst}) but instead  the general
$\left(1-2m_g/r+{q^2}/{r^2}\right)^{-1/2}$.

As discussed in Appendix 1, the differential inequality
(\ref{zineq}) implies that $\widetilde\z\le \widetilde\z_0$, for all
$f\le f(R)$, where $\widetilde\z_0(f)$ is the solution to the
differential equation ${d^2\widetilde\z_0}/{df^2}=
({Q^2}/{R^2})\widetilde\z_0$ satisfying the same boundary conditions
as $\widetilde\z$ does. One finds that
$\widetilde\z_0(f)=c_1^0e^{Qf/R}+c_2^0e^{-Qf/R}$, with
\bea c_1^0&\equiv&\frac12
e^{-\frac{Q}Rf(R)}\left(\frac{M}{Q}-\frac{Q}{R}+e^{-\L(R)}\right)\label{c0}\\
c_2^0&\equiv -&\frac12
e^{\frac{Q}Rf(R)}\left(\frac{M}{Q}-\frac{Q}{R}-e^{-\L(R)}\right).\nn\eea
Note that the conditions $\r_g' \sim (m_g/r^3)' \le 0$ and
$(q/r^3)'\ge 0$ imply that $e^{-2\L(r)}$ is always smaller than
$e^{-2\L_c}=(1-2Mr^2/R^3+Q^2r^4/R^6)$.  Therefore
$\widetilde\z_0(f(r)) \le \widetilde\z_c(f_c(r))$.

The conclusion is that $\widetilde\z(f(r))\le
\widetilde\z_0(f(r))\le \widetilde\z_c(f_c(r))$ or, equivalently
$\z(r)\le \z_c(r)$, where $\z(r)$ is the  general solution to
(\ref{zetadiff2}). Consequently, the critical radius for any
distribution with $\r_g'\le 0$, $q'\ge 0$ and $(q/r^3)'\ge 0$ is
always larger then the critical stability radius plotted in Fig.1.
The constant density case therefore provides us with an absolute
stability limit for any relativistic charged sphere satisfying these
conditions.

Of course, we do not know exactly what a ``realistic" charge
distribution is for such objects, and so below we provide a somewhat
cruder bound that is independent of any assumptions whatsoever about
the charge distribution.  The remainder of the paper consists of a
formal proof along the lines of YS, but we have streamlined the
presentation, filled in a number of gaps and present explicit
numerical results.

%-----------------------------------------------------------------------------
\section{General Case}
\setcounter{equation}{0}\label{sec7}
%-----------------------------------------------------------------------------

The plan is now to bound the behavior of the solution to
Eq.(\ref{zetadiff2}) for $0\le Q<M$ under the most general
conditions possible. To reiterate, we assume that any physically
acceptable solution meets only the following conditions:
\bea&& p(r) \ge 0 \;,\qquad \z(r) > 0\;,\nn\\
&&0\le Q< M\;,\qquad R> R_+ \;.\label{condition1}\eea

Here $R_+ \equiv M+\sqrt{M^2-Q^2}$ is the outer horizon of a RN
black hole; if $R = R_+$ then gravitational collapse of the charged
sphere has already taken place. The quantities $m_i$, $m_g$ and $q$
will be considered inputs that are related by Eq.(\ref{gravmassr})
and they satisfy the conditions
\bea&& m_g(R)=M\;,\qquad q(R)=Q\;,\nn\\
&& m_g\ge q\;,\quad m_g+\sqrt{m_g^2-q^2}<r\;.%,\quad m_i'\ge 0\;,\quad
%q'\ge 0\;,\quad \forall r\in[0,R).
\label{condition2}\eea
The last two conditions are required to avoid naked singularities,
as discussed in Section \ref{sec2}.

Once again, we define the {\it critical instability radius}
$R_c(M,Q)$ as the smallest radius $R > R_+$ for which a physically
acceptable solution can be found in $[0,R]$ {\it for {\it any} input
functions $m_g,q$} satisfying (\ref{condition2}).  We also
parameterize the difference between $Q$ and $M$ by
 \be
 \D \equiv \sqrt{1-Q^2/M^2} \; < 1 \label{Ddef}
 \ee
and assume $R = R_+(1+\e)$, where $\e$ is some number (not
necessarily small). With this notation, $R_+ = (1+\D)M$ and
 \be
 R = (1+ (1+\e)\D)M. \label{RD}
 \ee
We shall prove below that, under the conditions
(\ref{condition1})-(\ref{condition2}), the critical stability radius
admits a general lower bound of the form
\be R_c\ge \left(1+(1+\e_0)\D\right)M\label{uplowbounds}\ee
with $\e_0\simeq 1/264$. We do not believe that this specific value
of $\e_0$ has any physical relevance: it is merely a byproduct of
our estimates and it can certainly be improved.

Let us now turn to the proof of (\ref{uplowbounds}). The strategy
will consist in demonstrating that, whenever
$R<\left(1+(1+\e_0)\D\right)M$, for some suitable $\e_0$ to be
constructed below, then every possible solution $\z$ of Eq.
(\ref{zetadiff2}) becomes negative somewhere in $[0,R)$, for any
$m_g, q$ satisfying conditions (\ref{condition2}). Thus the solution
becomes physically unacceptable.  In the proof we shall need the
following preliminary estimate:\\

{\bf Lemma 1} {\it If $R<\left(1+(1+\e_0)\D\right)M$ and
$\d=(\e_0^2+2\e_0)^{1/2}$, then
\be \sqrt{1-\frac{2M}R+\frac{Q^2}{R^2}}\le \d
\left(\frac{M}{R}-\frac{Q^2}{R^2} \right)\label{est1}\ee}\\
For a proof of Lemma 1, see Appendix 2.

We now begin to study Eq.(\ref{zetadiff2}). Integrating both sides
between $r$ and $R$ and using the boundary conditions
(\ref{zetamatch}) we find, in analogy to Eq. (\ref{zeta'0}):
\be
\frac1{R^2}\left(\frac{M}{R}-\frac{Q^2}{R^2}\right)-\frac1re^{-\L(r)}\z'
(r)=\int_r^R dx\ e^{\L(x)}\left[
\left(\frac{m_g(x)}{x^3}\right)'-q(x)\left(\frac{q(x)}{x^4}\right)'\right]
\z(x)\label{int1}\ee
A crucial point in our proof consists in finding a uniform bound on
the right-hand-side, independent of $\z$, $m_g$ and $q$, at least
for $r$ close enough to $R$. This will allow us to dispense with the
details of $\zeta$. In Appendix 3 we prove the following key estimate.\\

{\bf Lemma 2} {\it Let $\b<1$. If $R<\left(1+(1+\e_0)\D\right)M$ and
$\e_0=\b^4\a_0$, with
\[
\a_0 =\frac{(\sqrt2-1)^2/4}{1+\sqrt{1+ (\sqrt2-1)^2/4}},
\]
then either the r.h.s. of (\ref{int1}) is uniformly bounded from
above by $\frac1{2R^2}\left(\frac{M}{R}-\frac{Q^2}{R^2}\right)$ for
all
$r\in[\b R,R]$ or $\z(r)=0$ for some $r\in[\b R,R)$.} \\

{\bf Remark.} Note that if $\e_0$ is chosen as in Lemma 2 then the
constant $\d$ in Lemma 1 satisfies $\d<\b^2(\sqrt2-1)/2$
(see the proof of Lemma 2).  This fact will be used below.\\

Now assume that $\z>0$ for any $r\le R$. By Lemma 2, for  any
$r\in[\b R,R]$ with $\b<1$, if $R<\left(1+(1+\e_0)\D\right)M$ where
now $\e_0=\b^4\a_0$, then the right hand side of (\ref{int1}) is
uniformly bounded above by $\frac1{2R^2}\left(\frac{M}{R}-
\frac{Q^2}{R^2}\right)$. As a consequence, from (\ref{int1}) we find
\[
\z'(r)\ge \frac{r e^{\L(r)}}{2R^2}\left(\frac{M}{R}-
\frac{Q^2}{R^2}\right).
\]
Integrating this inequality between $\b R$ and $R$ and using
$e^\L\ge 1$, we get:
\be \sqrt{1-\frac{2M}R+\frac{Q^2}{R^2}}-\z(\b R)\ge \frac12
\left(\frac{M}{R}-\frac{Q^2}{R^2}\right)\int_{\b R}^R \frac{ dr\,
r}{R^2},\ee
or
\be 0<\z(\b R)\le \sqrt{1-\frac{2M}R+\frac{Q^2}{R^2}}-\frac{1-\b}{4}
\left(\frac{M}{R}-\frac{Q^2}{R^2}\right)\ee
On the other hand, under the assumption that
$R<\left(1+(1+\e_0)\D\right)M$ with $\e_0=\b^4\a_0$, we have from
Lemma 1 and the remark after Lemma 2,
\[
0 > \sqrt{1-\frac{2M}R+\frac{Q^2}{R^2}} - \b^2\frac{
\sqrt2-1}2\left(\frac{M}{R}-\frac{Q^2}{R^2}\right).
\]
 If we merely set
 $\b^2\frac{ \sqrt2-1}2=\frac{1-\b}4$, which gives
 \[
 \b=\b_0\=\frac{\sqrt{1+8(\sqrt2-1)}-1}{4({\sqrt 2} - 1)}, %\cong .64999,
 \]
  we then have a
contradiction, and this implies that no physical solution to the
Einstein equations can be found. This proves (\ref{uplowbounds}),
with an explicit bound on $\e_0$ given by $\b_0^4\a_0$, or
$\e_0\simeq 1/264$.

We can now tabulate $R_c/M > 1 + (1+\e_0)\Delta$ for various values
of $Q/M$.  The results are shown in Table 2.

\vspace{2mm}
\begin{center}
\begin{tabular}{|c||c|}
\hline Q/M     &         $R_c$/M\\

0      &                2.250\\

.1      &                    1.999\\

.2       &                  1.983\\

.3       &                  1.956\\

.4       &                  1.920\\

.5      &                 1.869\\

.6       &                  1.830\\

.7       &                   1.717\\

.8       &                  1.602\\

.9       &                  1.437\\

.99       &                  1.142\\

.999      &                1.045\\

.9999     &                1.014\\
\hline
\end{tabular}\\
\vspace{3mm}

Table 2.  The lower bound on  $R_c/M$ is tabulated for and various
values of $Q/M$.
\end{center}

Because these figures represent a lower bound on $R_c/M$, they
should all lie beneath the corresponding numbers of Table 1, and
indeed they do.  {\it Any} relativistic charged sphere, regardless
of equation of state, must have a critical stability value of
$R_c/M$ greater than the values presented here. The discussion above
fails for $Q=M$, in which case the stability bound is simply
$R>R_+$.

%-----------------------------------------------------------------------------
\section{Conclusions}
\setcounter{equation}{0}\label{sec8}
%-----------------------------------------------------------------------------

The main result of this paper is an exact solution for the stability
limit of constant-density relativistic charged spheres for all $Q\le
M$. We also argued that in any ``physically reasonable" case where
the gravitational mass density decreases with the radius and the
charge density increases with radius, the constant-density case
provides an absolute stability limit for all charged spheres. If
calculation of a stability limit for rotating objects proves
tractable we would expect a quantitatively similar behavior for
$R_c/M$, given a value of $a/M$, the
angular-momentum-parameter-to-mass ratio.

For the most general charged case we found a cruder bound that is
independent of any assumption about the mass and charge
distribution, except for the basic conditions (\ref{condition2}).
Both ``physically reasonable" and general bounds approach the
horizon $R_+$ in the limit $Q\to M$.  That in this limit the
critical ``collapse'' radius is precisely $R_+$ is intuitively
reasonable because $Q=M$ is the point at which the Coulomb repulsion
equals the gravitational force. Such a state is evidently unstable,
but our results apparently  do not exclude what seems to be the only
route to producing an extremal black hole: to first create an
extremal charged sphere, then compress it to the horizon.
Nevertheless, given the other issues surrounding extremal bodies
mentioned in the Introduction, one should continue to hesitate
before regarding such objects as a smooth limit of the sub-extremal
state.

\\
{\bf Note Added:} Just prior to submission of this paper we learned
that a similar proof of our main result has been independently given in \cite{MDH01}.\\
\\
{\bf Acknowledgments.} The work of A.G. was partially supported by
U.S. National Science Foundation grant PHY 01 39984, which is
gratefully acknowledged.
\\

%-----------------------------------------------------------------------------
\section*{Appendix 1: On the differential inequality (\ref{zineq})}
\setcounter{equation}{0}
\renewcommand{\theequation}{A1.\arabic{equation}}
%-----------------------------------------------------------------------------

In this Appendix we prove that if $\widetilde\z(f)$ satisfies
inequality (\ref{zineq}) with boundary conditions
$\widetilde\z(f(R))=e^{-\L(R)}$ and $d\widetilde\z(f(R))/df
=M/R-Q^2/R^2$, then $\widetilde\z(f)\le \widetilde\z_0(f)$, for all
$0\le f\le f(R)$.  Again, $\widetilde\z_0(f)$ is the solution to the
differential equation $d^2\widetilde\z_0(f)/df^2 =(Q^2/R^2)
\widetilde\z_0(f)$ satisfying the same boundary conditions as
$\widetilde\z(f)$ does. It follows
 that the constant density case $m_g
\propto r^3, q \propto r^3$ provides an absolute stability limit for
any relativistic charged sphere in which $\r_g'\le 0$, $q' \ge 0$
and $(q/r^3)' \ge 0$.

In order to prove that $\widetilde\z\le\widetilde\z_0$, we define
$g(x)=\widetilde\z_0(f(R)-x)-\widetilde\z(f(R)-x)$ and  show that
$g(x)\ge 0$, for all $0\le x\le f(R)$. Note that $g(0)=dg(0)/dx=0$
and from (\ref{zineq}) that $d^2g/dx^2\ge (Q^2/R^2)\,g$, for all
$0\le x\le f(R)$. Without loss of generality, we can assume that
$g(x)$ is not identically zero in a right-neighborhood of the
origin. Let us pick some $x_0<R/Q$ (the reason for this choice will
become clearer below) and let $0\le x_1\le x_0$ satisfy
$g(x_1)=\min_{x\in[0,x_0]}g(x)$. There are two cases:
\\
\\
(1) $g(x_1)=0$. This means that $g(x)$ must be nonnegative in
$[0,x_0]$, which in turn implies that $g(x)$ is nonnegative for all
$0\le x\le f(R)$. If this were not the case, then there would be
some $x*$, for $0<x^*<f(R)$, such that $g(x^*)>0$ would be a local
maximum. But this would mean that $d^2g(x^*)/dx^2\le 0$, which
contradicts
 the condition $d^2g(x^*)/dx^2\ge (Q^2/R^2)\,g(x^*)>0$.
\\
\\
(2) $g(x_1)<0$. By the mean-value theorem $g(x_1)=x_1 dg(x_2)/dx$,
for some $0<x_2<x_1$. Again by the mean-value theorem,
$dg(x_2)/dx=x_2 d^2 g(x_3)/ dx^2$, for some $0<x_3<x_2$. On the
other hand $d^2g(x_3)/dx^2\ge (Q^2/R^2)\, g(x_3)$, and so we have
$0> g(x_1)\ge x_1 x_2 (Q^2/R^2)\, g(x_3)$, implying in particular
that $g(x_3)<0$. Since $g(x_1)$ is the minimum of $g(x)$ in
$[0,x_0]$ we also have $|g(x_3)|\le |g(x_1)|$.  Finally:
\be 0<|g(x_1)|\le x_1 x_2 (Q^2/R^2) |g(x_1)|\ee
Because $0<x_2<x_1\le x_0<R/Q$, we see that the r.h.s. of this
inequality is strictly smaller than $|g(x_1)|$, but this is a
contradiction. Thus $g(x_1)<0$ is an impossibility and the proof is
concluded.

%-----------------------------------------------------------------------------
\section*{Appendix 2: Proof of Lemma 1}
\setcounter{equation}{0}
\renewcommand{\theequation}{A2.\arabic{equation}}
%-----------------------------------------------------------------------------

Let $R= (1+ (1+\e)\D)M$ as in Eq. (\ref{RD}) for $0<\e<\e_0$. Then
in terms of $\D$ and $\e$:
\be
\frac{M}{R}=\frac1{1+(1+\e)\D}\;,\qquad\frac{Q^2}{R^2}=\frac{1-\D^2}{\left[
1+(1+\e)\D\right]^2}\ee
Inequality (\ref{est1}), which we want to prove, now takes the form:
\be 1-\frac2{1+(1+\e)\D}+\frac{1-\D^2}{\left[
1+(1+\e)\D\right]^2}\le \d^2
\left(\frac1{1+(1+\e)\D}-\frac{1-\D^2}{\left[
1+(1+\e)\D\right]^2}\right)^2.
 \ee
Multiplying both sides by $\left[1+(1+\e)\D\right]^2$ gives
\be \e^2+2\e \le \d^2\left(\frac{1+\e+\D}{1+(1+\e)\D}\right)^2.
\label{A2.2}\ee
Notice that for $0\le Q <M$ the right-hand-side is always $\ge
\d^2$. So if we choose $\d^2=\e_0^2+2\e_0$, Eq. (\ref{A2.2}) is
certainly satisfied and the lemma is proven. \qed

%-----------------------------------------------------------------------------
\section*{Appendix 3: Proof of Lemma 2}
\setcounter{equation}{0}
\renewcommand{\theequation}{A3.\arabic{equation}}
%-----------------------------------------------------------------------------

Let us denote the right-hand-side of (\ref{int1}) by $G(r)$. The
integral vanishes if the lower limit is $R$; hence $G(R)=0$. So, by
continuity, $G(r)$ will be less than $\frac1{2R^2}
\left(\frac{M}{R}-\frac{Q^2}{R^2}\right)$ in a small enough interval
of the form $[r_0,R]$. Let us pick some $\b<1$ and let
$\e_0=\b^4\a_0$, with $\a_0=\frac{(\sqrt2-1)^2/4}{1+\sqrt{1+
(\sqrt2-1)^2/4}}$ (the relevance of this specific number will be
made clear below). We want to show that either $G(r)\le
\frac1{2R^2}\left(\frac{M}{R}-\frac{Q^2}{R^2}\right)$ for all
$r\in[\b R,R]$, or $\z(r)=0$ for some $r\in[\b R,R)$, in which case
the solution becomes unphysical.

We proceed by contradiction. Assume that $\z(r)>0$ in $[\b R,R]$ and
that $G(r)-$ $\frac1{2R^2}$ $
\left(\frac{M}{R}-\frac{Q^2}{R^2}\right)$ changes sign in the same
interval. This means that there is some $\b'>\b$ such that both
$G(\b'R)=\frac1{2R^2}\left(\frac{M}{R}- \frac{Q^2}{R^2}\right)$ and
$G(r)<\frac1{2R^2}\left(\frac{M}{R}- \frac{Q^2}{R^2}\right)$,
$\forall r\in(\b'R,R]$. Now, from the definition of $e^{-\L}$ (Eq.
(\ref{eL})), one has
\be \frac{d(e^{-\L(x)})}{dx}=-x^2\,e^\L
\left[\left(\frac{m_g(x)}{x^3}\right)'-q(x)\left(\frac{q(x)}{x^4}\right)'
+2\frac{m_g}{x^4}-3\frac{q^2}{x^5}\right],\ee
 so that
$G(\b'R)$ can be immediately rewritten as:
\be G(\b'R)=\int_{\b'R}^R dx\
\left[-\frac1{x^2}\frac{d(e^{-\L})}{dx}\right]\z(x)+\int_{\b'R}^R
dx\ e^{\L(x)}\left(
-2\frac{m_g}{x^4}+3\frac{q^2}{x^5}\right)\z(x)\ee
By (\ref{condition2}), using in particular $q(x)\le m_g(x)\le x$, we
find that $-2{m_g}/{x^4}+3{q^2}/{x^5}\le 1/{x^3}$.  Thus, after
integrating the first term by parts,
\be G(\b'R)\le
-\frac{\z(R)}{R^2}e^{-\L(R)}+\frac{\z(\b'R)}{(\b'R)^2}
e^{-\L(\b'R)}+\int_{\b'R}^R dx\
e^{-\L}\left(\frac{\z'}{x^2}-2\frac{\z}{x^3} \right)+\int_{\b'R}^R
dx \frac{\z e^\L}{x^3}\label{uu}\ee
Note that from (\ref{press}), under the condition that $p\ge 0$ and
$r\ge m_g\ge q$, we must have $\z'\ge 0$ for $\z\ge 0$. So,
neglecting the negative terms in (\ref{uu}), and using the fact that
$e^{-\L}\le 1$, $\z' \ge 0 $ and that $x \ge \b' R$, the inequality
becomes:
\be G(\b'R)\le 2\frac{\z(R)}{(\b'R)^2} +
\frac{\z(R)}{(\b'R)^3}\int_{\b'R}^R dx\ e^{\L(x)}.\ee

Now, with the supposition that
$G(\b'R)=\frac1{2R^2}\left(\frac{M}{R}- \frac{Q^2}{R^2}\right)$ as
well as the boundary condition (\ref{zetamatch}) for $\z$,
\be \int_{\b'R}^R dx\ e^{\L(x)}\ge \frac{(\b')^3R}2\frac{\left(
\frac{M}R-\frac{Q^2}
{R^2}\right)}{\sqrt{1-\frac{2M}{R}+\frac{Q^2}{R^2}}}
-2\b'R\label{int2}\ee
On the other hand, in view of (\ref{int1}) and of the condition that
$G(r)<\frac1{2R^2}\left(\frac{M}{R}- \frac{Q^2}{R^2}\right)$,
$\forall r\in(\b'R,R]$, we have
\be \z'(r)> \frac{r e^{\L(r)}}{2R^2}\left(\frac{M}{R}-
\frac{Q^2}{R^2}\right)\;,\quad\forall r\in(\b'R,R].\ee
Integrating between $\b'R$ and $R$ yields
\be \z(R)-\z(\b'R) > \frac1{2R^2}\left(\frac{M}{R}-
\frac{Q^2}{R^2}\right)\int_{\b'R}^R dr\, r e^{\L(r)},\ee
and employing the boundary conditions on $\zeta(R)$ once again gives
\be \z(\b'R)<\sqrt{1-\frac{2M}R+\frac{Q^2}{R^2}}-\frac{\b'}{2R}
\left(\frac{M}{R}-\frac{Q^2}{R^2}\right)\int_{\b'R}^R dr \,
e^{\L(r)}.\ee
Since inequality (\ref{int2}) gives a minimum for the integral in
this expression we can insert (\ref{int2}) into into the
right-hand-side here to get, finally,
\be 0<\z(\b'R)<\sqrt{1-\frac{2M}R+\frac{Q^2}{R^2}}-\frac{\b'}{2R}
\left(\frac{M}{R}-\frac{Q^2}{R^2}\right)\left[
\frac{(\b')^3R}2\frac{\left( \frac{M}R-\frac{Q^2}
{R^2}\right)}{\sqrt{1-\frac{2M}{R}+\frac{Q^2}{R^2}}}
-2\b'R\right].\ee
If we let $y=\sqrt{1-\frac{2M}{R}+\frac{Q^2}{R^2}}/\left(
\frac{M}R-\frac{Q^2}{R^2}\right)$, then the previous expression
becomes
\be y^2+(\b')^2y-\frac14(\b')^4>0, \ee
which implies that
\be y>\b'^2\frac{\sqrt2-1 }2 > \b^2\frac{\sqrt2-1 }2.
 \ee
On the other hand, we see that Lemma 1 states that $y \le \d$, where
$\d=(\e_0^2+2\e_0)^{1/2}$. Take $\e_0$ smaller than the positive
root of the equation $x^2+2x-\b^4(\sqrt2-1)^2/4=0$, that is, smaller
than $x_+= -1+\sqrt{1+ \b^4(\sqrt2-1)^2/4}$.  For example,
$\e_0=\b^4/47.13 < \b^4 (\sqrt2-1)^2/[4+2\sqrt{4+ (\sqrt2-1)^2}]$
does the job. Then $y<\b^2\frac{\sqrt2-1}2$. We have reached a
contradiction, and the Lemma is proved.\qed

\end{document}